\documentclass[]{spie}  %>>> use for US letter paper
\pdfoutput=1
\usepackage{amsmath,amsfonts,amssymb}
\usepackage{graphicx}
\usepackage{epstopdf}

\usepackage{selinput}
\SelectInputMappings{
  adieresis={ä},
  germandbls={ß},
  Euro={€}
}

\newcommand{\keyword}[1]
           {
             \noindent
             {\bf{\em #1}} \hspace*{2mm}
           }

\newcommand{\simless}{\mathbin{\lower 3pt\hbox
     {$\rlap{\raise 5pt\hbox{$\char'074$}}\mathchar"7218$}}}
\newcommand{\simgreat}{\mathbin{\lower 3pt\hbox
     {$\rlap{\raise 5pt\hbox{$\char'076$}}\mathchar"7218$}}}
\newcommand{\about}    {$\sim$\ }

\usepackage{amsmath,amsfonts,amssymb}
\usepackage{graphicx}
\newcommand{\mum}{$\mu$m}
\newcommand{\farcs}{\mbox{\ensuremath{.\!\!^{\prime\prime}}}}

\newcommand{\AU}{au}
\newcommand{\HD}{HD\,}

\title{Science with MATISSE}

\author[a]{Sebastian Wolf}
\author[b]{Bruno Lopez}
\author[c]{Jean-Charles Augereau}
\author[b]{Marco Delbo}
\author[d]{Carsten Dominik}
\author[e]{Thomas Henning}
\author[f]{Karl-Heinz Hofmann}
\author[g]{Michiel Hogerheijde}
\author[h]{Josef Hron}
\author[g]{Walter Jaffe}
\author[b]{Thierry Lanz}
\author[e]{Klaus Meisenheimer}
\author[b]{Florentin Millour}
\author[i]{Eric Pantin}
\author[b]{Roman Petrov}
\author[f]{Dieter Schertl}
\author[e]{Roy van Boekel}
\author[f]{Gerd Weigelt}
\author[b]{Andrea Chiavassa}
\author[j]{Attila Juhasz}
\author[b]{Alexis Matter}
\author[b]{Anthony Meilland}
\author[b]{Nicolas Nardetto}
\author[k]{Claudia Paladini}

\affil[a]{Universität zu Kiel, Institut für Theoretische Physik und Astrophysik, Leibnizstr. 15, 24098 Kiel, Germany}
\affil[b]{Laboratoire Lagrange, UMR7293, Universit\'{e} de Nice Sophia-Antipolis, CNRS, Observatoire de la C\^ote d'Azur, Nice, France}
\affil[c]{UJF-Grenoble 1/CNRS-INSU, Institut de Plan\'{e}tologie d'Astrophysique de Grenoble (IPAG) UMR 5274, Grenoble, 38041, France}
\affil[d]{Sterrenkundig Instituut "Anton Pannekoek", Science Park 904, 1098 XH, Amsterdam, The Netherlands; Afdeling Sterrenkunde, Radboud Universiteit Nijmegen, Postbus 9010, 6500 GL, Nijmegen, The Netherlands}
\affil[e]{Max-Planck-Institut für Astronomie, Königstuhl 17, 69117 Heidelberg, Germany}
\affil[f]{Max-Planck-Institut für Radioastronomie, Auf dem Hügel 69, 53121 Bonn, Germany}
\affil[g]{Sterrewacht Leiden, Universiteit Leiden, Niels-Bohr-Weg 2, 2300 CA, Leiden, The Netherlands}
\affil[h]{Institut für Astronomie, Universität Wien, Türkenschanzstraße 17, 1180 Wien, Austria}
\affil[i]{CEA/DSM/IRFU/Service d'Astrophysique, CE Saclay, France}
\affil[j]{Institute of Astronomy, University of Cambridge, Madingley Road, Cambridge, CB3 0HA, United Kingdom}
\affil[k]{Institut d’Astronomie et d’Astrophysique, Universite’ libre de Bruxelles, Boulevard du Triomphe CP 226, B-1050 Bruxelles, Belgium}

\authorinfo{Further author information: Send correspondence to S. Wolf (wolf@astrophysik.uni-kiel.de)}

\begin{document} 
\maketitle

%%%%%%%%%%%%%%%%%%%%%%%%%%%%%%%%%%%%%%%%%%%%%
% abstract
\begin{abstract}
We present an overview of the scientific potential of 
MATISSE, the Multi Aperture mid-Infrared SpectroScopic Experiment for the Very Large Telescope Interferometer.  
For this purpose we outline selected case studies from various areas, such as star and planet formation, active galactic nuclei, evolved stars, extrasolar planets, and solar system minor bodies
and discuss strategies for the planning and analysis of future MATISSE observations.
Moreover, the importance of MATISSE observations in the context of complementary high-angular resolution observations at near-infrared and submillimeter/millimeter wavelengths is highlighted.
\end{abstract}

\keywords{Astrophysics, Long-baseline interferometry, Infrared, Very Large Telescope Interferometer, MATISSE}

%%%%%%%%%%%%%%%%%%%%%%%%%%%%%%%%%%%%%%%%%%%%%
% body

% ==========================================================
% Intro
\section{Introduction} \label{sect:intro} 
MATISSE, the Multi Aperture Mid-Infrared SpectroScopic Experiment\cite{2014Msngr.157....5L}, is foreseen as a mid-infrared spectro-interferometer combining the beams of up to four Unit telescops (UTs) or Auxiliary telescopes (ATs) of the Very Large Telescope Interferometer (VLTI). 
MATISSE will measure closure phase relations and thus offer an efficient capability for image reconstruction. 
MATISSE will open two new observing windows at the VLTI: The L and M bands in addition to the N band with the possibility to perform simultaneous observations in separate bands. 
MATISSE will be equipped with several spectroscopic modes with spectral resolutions up to
\about\,5000. These modes will provide unique insight in the physical properties and kinematics of line emitting gas regions, for example in the environment of young stellar objects (YSOs) and hot stars.

MATISSE will extend the astrophysical potential of the VLTI in the mid-infrared wavelength range by overcoming the ambiguities often existing in the interpretation of simple visibility measurements. The existence of the four large apertures of the VLTI will permit to push the sensitivity limits up to values required by selected astrophysical programs such as the study of Active Galactic Nuclei and protoplanetary disks around low-mass stars. Moreover, the existence of the ATs which are relocatable in position will allow the exploration of the Fourier plane with up to 200 meters baseline length. Key science programs using the ATs cover for example the formation and evolution of planetary systems, the birth of massive stars as well as the observation of the high-contrast environment of hot and evolved stars.

In most astrophysical domains which require a multi-wavelength approach, MATISSE will be a perfect complement to other high angular resolution facilities. MATISSE covers the mid-infrared spectral domain, between the near-infrared domain, for which various high-angular resolution instruments are available (e.g., SPHERE/VLT, PIONIER/VLTI) and submillimeter/millimeter wavelengths at which ALMA, the Atacama Large Millimeter Array, operates. 
With the extended wavelength coverage from the L to the N band, MATISSE will not only allow one to trace different spatial regions of the targeted objects, 
but also different physical conditions and processes as well as 
the possibly varying origin of the radiation (thermal radiation vs.\ scattered light) and 
thus provide insights into previously unexplored areas, 
such as the investigation of the distribution of volatiles in addition to that of the dust.

In the following science cases for MATISSE 
from the fields of 
star and planet formation (Sect.~\ref{sec:psf}),
stellar physics  (Sect.~\ref{sec:stellphys}),
planetary systems (Sect.~\ref{sec:planetsys}), and
Active Galactic Nuclei (Sec.~\ref{sec:agn})
are presented.

% ==========================================================
% ``Science body''
\section{Physics of star and planet formation}\label{sec:psf}

% Protoplanetary disks: Dust phase
\subsection{Class II YSOs / gas-rich disks}
\label{sec:class_II}  % \label{} allows reference to this section

%\subsubsection{Introduction}
%\label{sec:class_II:introduction}

The circumstellar disks around newly-formed stars have been subject to intense study for decades. Their structure and chemical composition are interesting because they are the birth places of new planetary systems. Determining the properties of the ``building material'' for planets, as well as measuring signatures of interaction between planets and the disk material, provide much needed observational constraints on models of disk evolution and planet formation. However, the small angular scales involved pose technological challenges: at a distance of 150\,pc, typical for nearby young stars, the orbital radius of Jupiter corresponds to only 0\farcs035. This makes direct study of the disks at these scales the realm of interferometers. The innermost $\approx$15~\AU \ of the disks are the most interesting in terms of planet formation on solar system scales. These can be well observed in the near- and thermal infrared where we predominantly trace the warm surface layer of the disks.
For an overview of science cases for next-generation optical/infrared long-baseline interferometers 
in the field of circumstellar disks and planets see\cite{2012A&ARv..20...52W}.

In this section we present some of the capabilities of MATISSE for studying disks of ``intermediate'' age that are commonly referred to as ``class~II'' objects\cite{1987IAUS..115....1L}. They are typically 1 to a few million years old, and contrary to the younger ``class~I'' objects they are no longer deeply embedded in their natal clouds. The class~II disks are still relatively massive and gas-rich, with typical disk/star mass ratios of order $10^{-2}$ and at least an order of magnitude scatter around this value \cite{2011ARA&A..49...67W}. Most T\,Tauri and Herbig~Ae (HAe) stars are class~II objects. They are thought to be the predecessors of the debris disk objects that have lost their primordial gas and dust and have much less massive disks of secondary dust, produced in collisional grind-down of larger bodies. The class~II phase lasts a few million years \cite{2010A&A...510A..72F}, during which there is typically still substantial accretion onto the central star at rates approximately following the relation $\dot M \approx 10^{-8} (M_*/{\rm M}_{\odot})^2$, where $\dot M$ is the accretion rate in ${\rm M}_{\odot}$/yr and $M_*$ denotes the stellar mass, with about two orders of magnitude scatter about this average relation \cite{2006A&A...452..245N,2008ApJ...681..594H,2013ApJS..207....5F}. The class~II phase is thought to be the main phase of planet formation in the ``core accretion paradigm'' \cite{1986Icar...67..391B}, and at least gas giant planet formation must be completed before the gaseous disk has dissipated.

The spatial structure of the disks around T\,Tauri and HAe stars has been intensively studied in order to constrain the evolution of the disks that finally leads to their dissipation, and to search for signatures of forming planets and their interaction with the disk material. Because the angular extent of the disk emission is small, this has been the realm of interferometers. At different wavelengths we see physically different parts of the disk. In the near-infrared we can observe thermal emission from material in the very inner disk, typically at less than 1~\AU \ from the star. Around 10\,\mum \ we see a larger region of order 10~\AU \ in size for a typical HAe star\footnote{The thermal emission of disks arises from dust grains that are in equilibrium with the radiation field. The spatial extent of the region we see at a given wavelength is that where the temperature is high enough for the Planck function to not be deeply in the Wien regime. The size of this region scales approximately with $\sqrt{L_*} \lambda^2$, where $L_*$ is the stellar luminosity and $\lambda$ the observing wavelength.} that originates in the warm disk surface layer. At millimeter wavelengths we mostly see the cooler disk interior, and can observe emission out to $>$100~\AU, provided the disk has an appreciable optical depth at these large radii. Typical apparent sizes for a disk around a HAe star at a distance of 150~pc are $<$10~mas for the thermal emission in the H and K bands, $\approx$0\farcs1 at 10~\mum, and up to a few arcseconds for the emission at millimeter wavelengths\cite{2003ApJ...596..597W}. In addition to the thermal disk emission we may observe stellar light scattered off dust grains in the disk surface out to $>$100~\AU\cite{2007ApJ...665.1391G,2013ApJ...771...45D}, and in some cases we may observe emission from stochastically heated particles like Polycyclic Aromatic Hydrocarbons (PAHs)\cite{2007A&A...470..625D}.

It has become apparent that the class~II disks show much structure, such as gaps and rings in the large grain distribution near the midplane as seen in millimeter observations\cite{2015ApJ...808L...3A,2016arXiv160309352A,2016arXiv160303731C} and in scattered light\cite{2015ApJ...802L..17A,2015ApJ...815L..26R}, and large scale spirals seen in scattered light observations \cite{2015A&A...578L...6B,2016arXiv160300481S}. In the innermost few \AU \ of several disks evidence for gaps was found using infrared interferometry, in particular combining near-infrared and 10~\mum \ observations \cite{2012A&A...541A.104C,2013A&A...555A.103S,2014A&A...561A..26M}.

%Class II and Class III objects, refer to jean-Charles piece. Gas rich disks, transition disk phase, connection with planet formation. 

%the MIDI heritage.

%MATISSE: more baselines, more wavelengths, and images.

\subsubsection{The MIDI heritage}
\label{sec:classII:MIDI_heritage}

The MIDI instrument at the VLT Interferometer\cite{2000SPIE.4006...43L} was the predecessor of MATISSE operating at 10~\mum. It was a two-element optical interferometer, capable of spectrally resolved measurements between 8 and 13~\mum. It was the first facility that allowed systematic investigation of disks at these wavelengths with a spatial resolution of $\lesssim$10~mas, corresponding to a linear resolution close to 1~\AU \ at a distance of 150~pc. MIDI saw first light in December 2002 and in the 12 years of regular observations that followed well over 100 disks were studied. These experiments resulted in numerous new insights, such as a correlation between spatial extent of the disk emission and SED shape \cite{2004A&A...423..537L}, radial gradients in dust composition \cite{2004Natur.432..479V,Schegerer:2008kf}, the detection of gaps \cite{2012A&A...541A.104C,2013A&A...555A.103S,2014A&A...561A..26M}, and the detailed measurement of the ``wall'' shape in transition disks \cite{2013A&A...557A..68M,2014A&A...564A..93M}. 

\begin{figure}[t]
\begin{center}
\includegraphics[height=8cm]{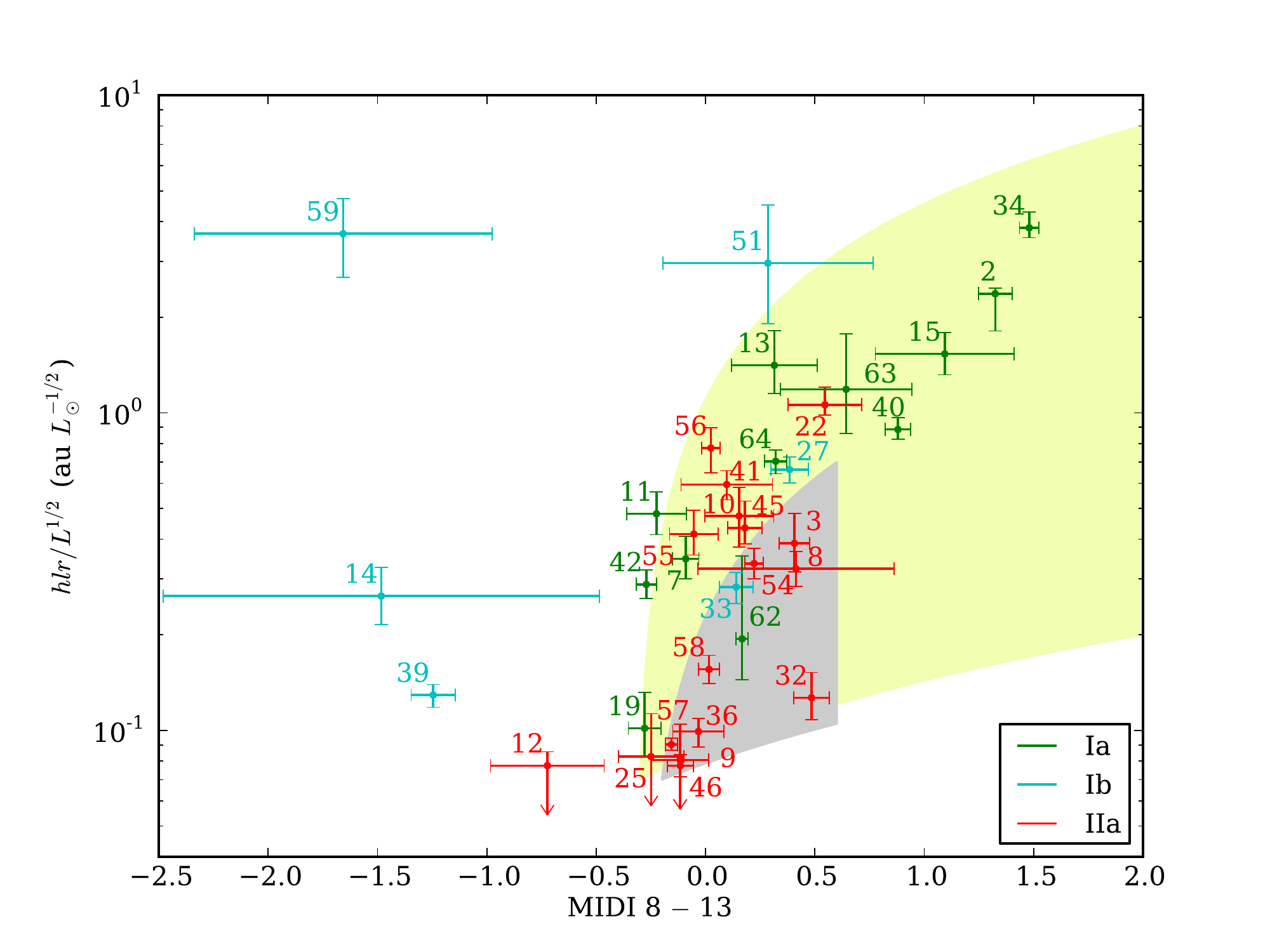}
\end{center}
\caption[MIDI survey of HAe stars]{\label{fig:fig_RvB_1}The MIDI size-color diagram for Herbig Ae star disks (figure adopted from \cite{2015A&A...581A.107M}).}
\end{figure}

The largest and most systematic analysis of the disk observations performed with MIDI was recently presented\cite{2015A&A...581A.107M}. The key results of this study are summarized in Fig.~\ref{fig:fig_RvB_1}, where the MIDI size-color diagram of the studied sample of HAe star disks is shown. In the vertical direction sources go from compact to extended (bottom to top), in the horizontal direction from blue to red (left to right). The ``shark fin-shaped'' regions show the ranges occupied by radiative transfer models of disks\cite{2015A&A...581A.107M}. The grey region shows the range occupied by continious, gap-less models. The larger yellow region shows where radiative transfer models with a range of gap sizes lie. It was found that: 
\begin{enumerate}
\item Sources group~I-type\footnote{Note that the group~I vs. group~II classification\cite{2001A&A...365..476M} is an SED-based sub-division of the optically visible, gas-rich disks. Both group~I and group~II sources are class~II\cite{1987IAUS..115....1L} objects.} SEDs\cite{2001A&A...365..476M} (traditionally associated with flared disk geometries, green symbols in Fig.~\ref{fig:fig_RvB_1}) show evidence of large gaps in the spatially resolved data, confirming an earlier hypothesis; 
\item Many sources with group~II-type SEDs (traditionally thought to have flat disks, red symbols in Fig.~\ref{fig:fig_RvB_1}) have the signature of a continuous, ``gap-less'' disk, but 
\item Some group~II sources do show evidence for (smaller) gaps in their inner disk regions.
\end{enumerate}

The emergent picture is that the signatures of radial gaps are much more common among HAe star disks than previously thought. Should these gaps be caused by planet-disk interaction, then planets of $\approx$1 Saturn mass or more must be very common in these young disks, opening exciting prospects of detecting and characterizing them with e.g. METIS\cite{2010SPIE.7735E..2GB} at the E-ELT. In particular, the discovery of group~II sources with gaps is exciting; these gaps are too small to be noticable in the SED and could only be uncovered with the high spatial resolution that MIDI offered. With MIDI we could demonstrate the presence of gaps in the inner disks of many objects. With MATISSE we can take the next step and accurately measure the gap profiles and possibly detect small-scale asymmetries.

\subsubsection{The future: MATISSE}
\label{sec:classII:MATISSE}

The MIDI data revealed a large diversity of inner disk geometries. However, due to the limited capabilities of MIDI (1 simultaneous baseline, no closure phases, only N band), the geometry of the inner disks and gaps could be studied in little detail only. Taking the next step requires the qualitatively new capabilities of MATISSE\cite{2014Msngr.157....5L}. With its four-element beam combiner it delivers 6 baselines as well as 3 independent closure phases with each individual measurement. This yields qualitatively new observational possibilities:
\begin{itemize}
\setlength{\itemsep}{-0.001cm}
\item Precise mapping of the radial intensity profiles, localizing gaps and measuring their shapes,
\item Measurement of inner disk inclinations and position angles,
\item Sensitive measurement of deviations from point symmetry using closure phases,
\item Higher resolution mapping of the inner $\approx$\,1~\AU \ using the 3-5~\mum \ spectral range,
\item Measurement of the gas kinematics in the inner disk and accretion columns (see Sect.~\ref{ssec:lines}), and
\item Model-independent image re-construction.
\end{itemize}

\keyword{Exquisite survey power}
The inner disks can be surveyed for signatures of gaps or asymmetries with a single or few measurements per source, allowing to efficiently identify objects exhibiting sub-structure, which can then be followed-up with better UV coverage and be mapped in detail. The large wavelength range (3-14~\mum) allows probing the disk from $\lesssim$1~\AU\ to $\approx$10~\AU. The availability of an external fringe tracker is essential for surveys because it allows larger samples of objects to be observed with the ATs.

%\emph{Write a paragraph about surveys?}
\keyword{Model-independent image re-construction}
%\subsubsection{Model-independent image re-construction}
%\label{sec:classII:image_reconstruction}
%
\begin{figure}[t]
\begin{center}
\begin{tabular}{ccc}
\hspace{-0.5cm}
\includegraphics[width=0.323\textwidth,trim=0 0 0 0, clip]{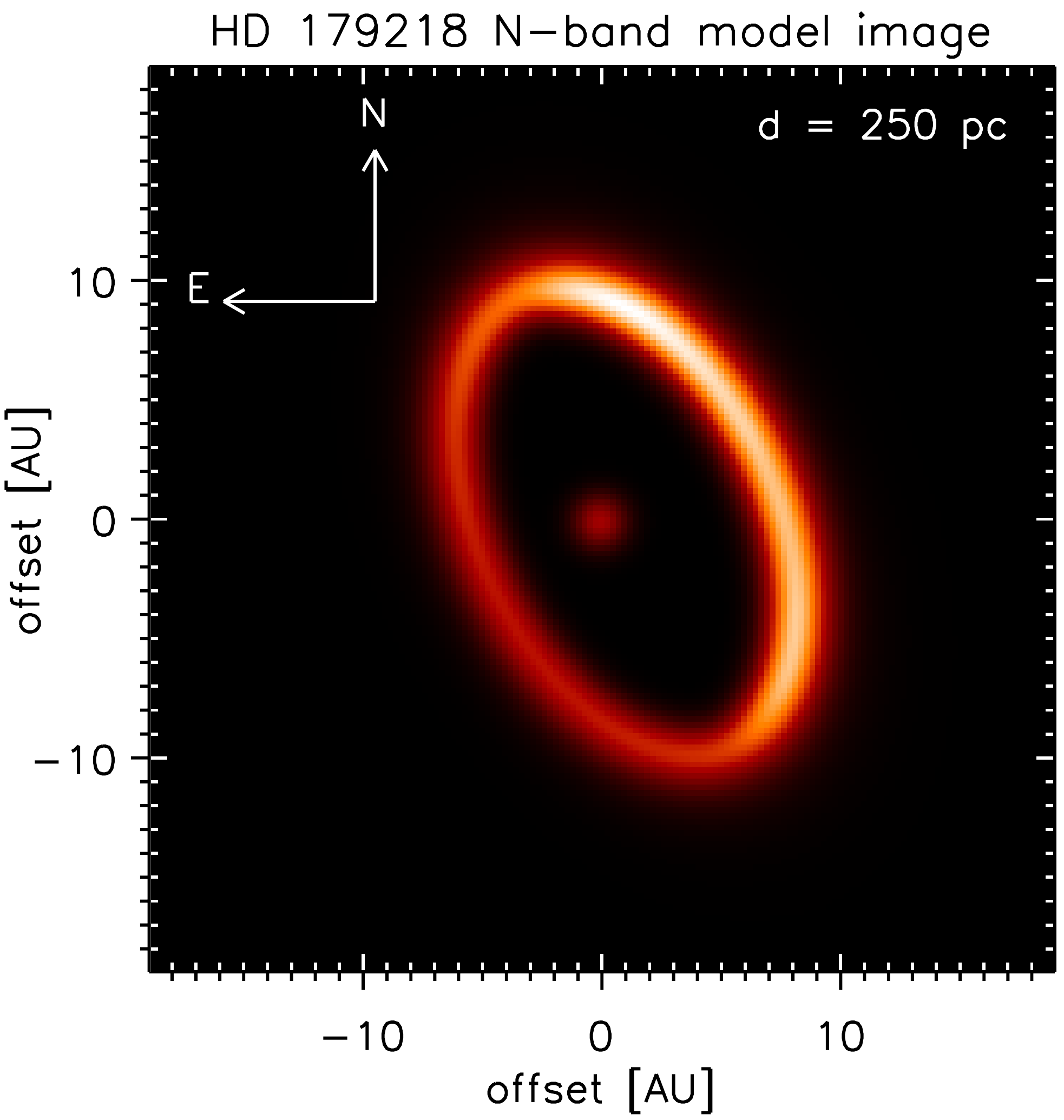} &
\includegraphics[width=0.323\textwidth,trim=0 0 0 0, clip]{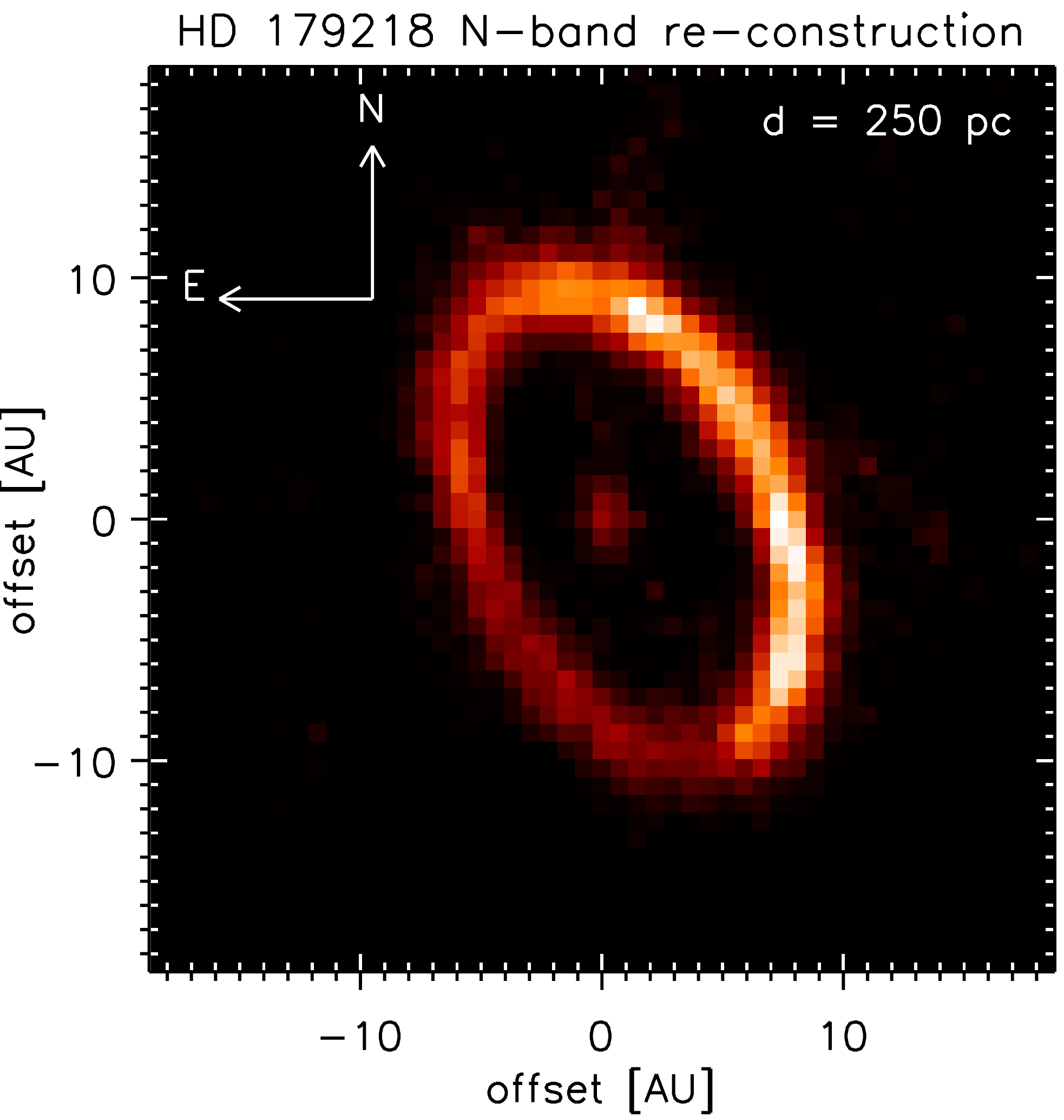} &
\includegraphics[width=0.34\textwidth,trim=385 0 0 0, clip]{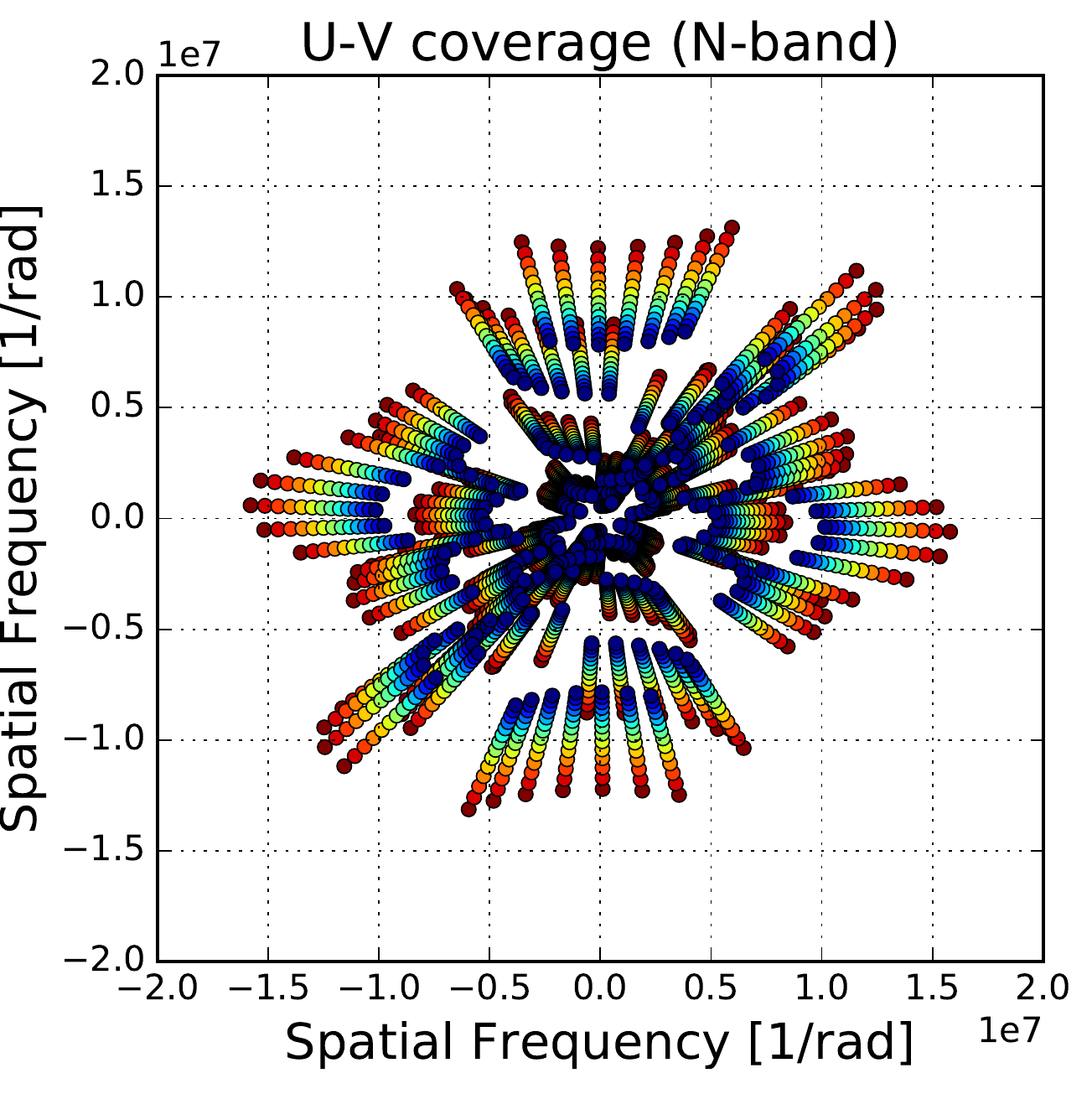}
\end{tabular}
\end{center}
\caption[HD~179218 image re-construction]{\label{fig:HD179218}An image re-construction experiment using simulated MATISSE data of the transition disk \HD179218. \emph{Left}: radiative transfer model image; \emph{center}: re-constructed MATISSE image; \emph{right}: UV coverage.}
\end{figure} 
MATISSE was designed to allow model-independent image re-construction by measuring visibility amplitudes and closure phases at a broad range of different baselines. The re-configurable ATs are essential to sample both the low spatial frequencies with a compact configuration as well as the high spatial fequencies on longer baselines. Also here, an external fringe tracker is essential for image re-construction experiments because the required UV coverage can only be achieved with the ATs.

In Fig.~\ref{fig:HD179218} we show an example of a MATISSE image re-construction experiment, targeted at the transition disk object \HD179218. The left panel shows a radiative transfer model that provides a good fit to a set of 18 MIDI observations (Menu et al., subm.). The MIDI data indicate a large gap with a radius of 10~\AU, a small amount of spatially unresolved warm dust around the star, and a slightly eccentric gap shape. 
The middle panel shows an image that was re-constructed from simulated MATISSE observations
using the software SQUEEZE\footnote{https://github.com/fabienbaron/squeeze}, on 3 AT configurations and with 7 measurements per 
configuration\footnote{The re-constructed image was obtained using a pixel scale of 3~mas. The L0-norm was used as regularizer and the value of the hyper parameter $\mu$ was set to 10000. No a-priori assumption on the target’s brightness distribution was made. The resulting image that fits the simulated V$^2$ and closure phases with a total $\chi^2$=1.5.}. 
In this experiment an average N band image was made from a series of monochromatic images in the 8-13~\mum \ range, and data in the whole wavelength range were used together in the re-construction of a single, average image. Methods for simultaneous multi-wavelength re-construction are being developed and should in time replace this simplified approach.

% Protoplanetary disks: Gas phase
\subsection{Observing molecular line radiation from protoplanetary disks}
\label{sec:lines}

MATISSE not only promises a rich discovery space for observations of
the thermal emission of dust grains in planet-forming disks, but is
also capable of detecting spectral lines at resolutions up to 3500
(M-band) to 5100 (L-band). Corresponding to velocity resolutions of
$<$100 km s$^{-1}$, such observations promise to reveal key
astrophysical processes related to the forming star: Atomic lines such
as Br$\gamma$ and Pf$\beta$ trace accretion from the disk onto the
star as well as high velocity winds and jets driven by the star
\cite{kraus2008, weigelt2011, caratti2015, garcia2015}. The
spatial resolution of MATISSE of 0.5--1.0 au in typical nearby star
forming regions provides strong constraints on geometry of accretion
and the launch points of the winds and jet.

In the M-band, MATISSE offers the unique capability to obtain high
angular resolution observations of the ro-vibrational ground state
lines of CO, around 4.67 $\mu$m. The lines, distributed over a $P$ and
$R$ branch, are separated by $\sim$0.04 $\mu$m (2500 km s$^{-1}$). At
a resolution R$\sim$3500 MATISSE can easily separate the individual
rotational lines. High velocity (tens to a few hundred km s$^{-1}$)
emission of CO is expected to trace the disk winds if they are
launched near the inner edge of the disk. Together with the atomic
lines described above, MATISSE is likely to offer important insight
into the wind/jet driving mechanism, which is an essential process in
the star formation that regulates the angular momentum of the
accreting star.

The CO ro-vibrational ground state lines have been extensively studied
at very high spectral resolution ($R\sim$100,000) using long-slit
spectroscopy with VLT/CRIRES and Keck/NIRSPEC \cite{brittain2003,
  blake2004, rettig2004, salyk2007, salyk2011, pontoppidan2011}. These
observations do not spatially resolve the emission, but from the
resolved line profiles the emission is found to originate from the
inner 0.1 au of the disk. The occurence of these lines suggests that
the inner disks, even if cleared out of dust, still hold (some)
gas. The latter explains the continued accretion that is observed onto
these stars.  However, the mechanism through which gas can filter
through the inner disk holes is unknown. Does gas uniformly fill the
hole, suggesting that (some) gas can filter past the inner dust disk
edge? Or is gas subject to dynamical interaction with a
planet/companion in the hole, forming a spiral accretion flow onto the
star (or even onto the planet)?

\begin{figure} [t]
  \begin{center}
%    \begin{tabular}{c} %% tabular useful for creating an array of images 
      \includegraphics[height=6cm]{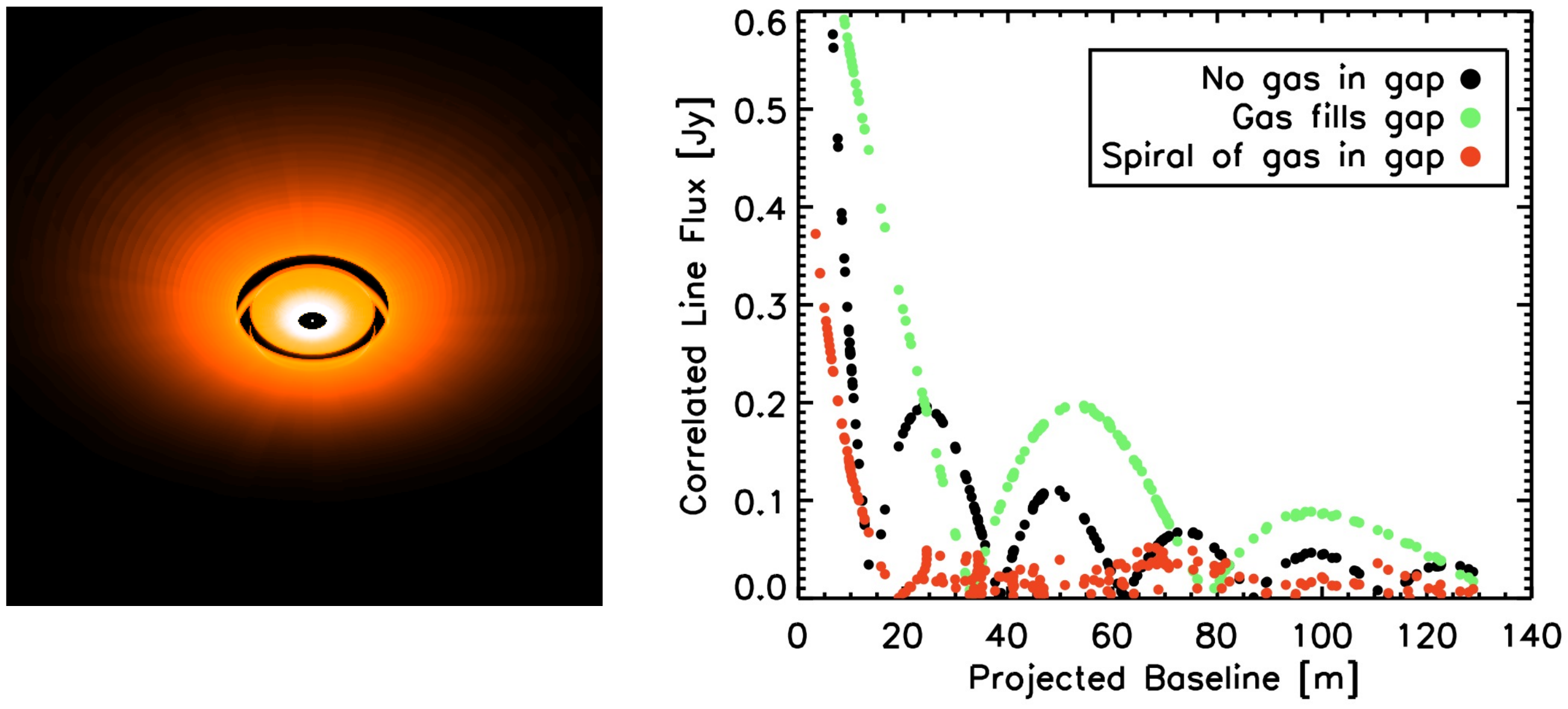}
%    \end{tabular}
  \end{center}
  \caption[example] 
  {Left: Simulated image of a Herbig Ae disk with a cleared out gap,
    seen under an inclination of 45$^\circ$. Right: Predicted
    correlated flux in CO-line emission channels for three different
    models: with no gas in the gap (black symbols), with the gap
    uniformly filled with gas (green symbols), and with a spiral
    streamer of gas in the otherwise empty gap (red
    symbols).  \label{fig:disks_co_simulation}}
\end{figure}

The MATISSE M-band resolution of 3500 will allow us to seperate the
individual rotational lines of the CO P- and R-branches of the
ro-vibrational ground-state spectrum from the intervening
continuum. MATISSE will not velocity-resolve the individual lines,
which NIRSPEC and CRIRES observations show to extend to 100 km
s$^{-1}$ or less. By studying the differential visibilities between
the continuum and the CO lines, we will be able to investigate if (a)
a hole is present in CO just like the hole seen in continuum, or (b)
the gas uniformly fills the hole. Furthermore, the closure phases
provided by MATISSE will immediately tell us if the CO emission is
symmetric or confined to asymmetric strcutures such as spiral
arms. Figure \ref{fig:disks_co_simulation} illustrates the three
different scenarios.

To assess the feasibility of using MATISSE to study the CO
ro-vibrational emission from disks, we performed simulations based on
simple physical descriptions of the disk and calculated synthetic
visibilities. These calculations show that MATISSE has sufficient
sensitivity to detect the CO visibilities for a handful of bright
Herbig systems, using the UT telescopes. The AT telescopes lack
collecting area for this work, and the lower mass (and less bright) T
Tauri disks are too weak to be detected even on the UTs. However, a
small sample of Herbigs with disks that are very well studied (and
that will most certainly be the subject of continuum work with MATISSE
as well) will allow us to address the relative distribution of dust
and molecular gas in the inner disk of these representative systems.

Absorption by atmospheric CO is a major complication in these MATISSE
observations. The very high resolution CRIRES and NIRSPEC observations
described above use the Doppler shift of the source at different
epochs to remove the atmospheric absorption from the source
emission. At the lower spectral resolution of MATISSE, source emission
and atmospheric absorption always coincide in the same spectral
resolution element. A careful observational setup is therefore
required, that also includes observing a strong calibration source
with a featureless spectrum, to characterize the atmospheric
absorption. Since the atmospheric absorption occurs in a smooth
screen, the relative absorption depth follows from the differential
visibilities between the pure-continuum and line-containing spectral
channels obtained toward the calibrator. The additional differential
visibilities between pure-continuum and line-containing spectral
channels toward the source correspond to real structure in the CO
emission lines. With detailed model calculations based on the known
(spectrally resolved, spatially unresolved long-slit) spectra,
specific spatial distributions of CO can then be confirmed or
rejected, such as gas filling the gap uniformly or in a spiral arm.

With its unprecedented combination of angular and spectral resolution,
MATISSE will offer important new insight on the relative distributions
of dust and gas in the inner disk regions. With MATISSE we can answer
important questions, such as how gas accretes from the inner disk edge
onto the star and/or possible (planetary?) companions and what the
launching region and mechanism is of the powerful jets and winds
driven by forming stars.

\label{ssec:lines}

% ----------------------------------------------------------------------------------------------------------------------------
\section{Stellar physics}\label{sec:stellphys}
Besides the investigation of the gas/dust phase of protoplanetary disks,
MATISSE will also offer the rare opportunity to image the dust around stars, right at the location where it forms. This will allow one to put constraints on the circumstellar structure, on the mass ejection and reorganization, and on the nature and formation processes of dust. 
MATISSE measurements will be pivotal for the understanding of large multiwavelength datasets on the same targets collected through various high-angular resolution and high-spectral resolution facilities at ESO, covering the infrared to millimeter wavelength range 
(e.g., 
near-infrared adaptive optics: NACO, SPHERE;
long-baseline interferometry in the near-infrared: PIONIER, GRAVITY;
near/mid-infrared spectroscopy: CRIRES;
mid-infrared imaging: VISIR;
submillimeter/millimeter interferometry: ALMA).
Among main-sequence and evolved stars, several cases of interest have been identified that are described briefly in the following (see also\cite{2015ASPC..497.....K,2015Msngr.161...43W} for an overview of the most important scientific results and open questions):

\subsection{Massive stars}

Stars with masses $>8\,{\rm M}_{\odot}$ evolve fast and blast as supernovae at the end of their lives, not without having gone through a firework of different physical processes.
Here, we will focus on a selected cases of the various kinds of massive stars:

%It is not the purpose of this part to describe all different kind of massive stars 
%\citep[see][]{todo{Millour et al. 2016, 
%Weigelt et al. 2016, 
%Soulain et al. 2016}}
%and we will focus on a few selected cases for this paper.

\keyword{Dusty blue supergiant stars}
Several blue supergiant stars produce large amounts of dust before exploding as supernovae. Though they are not the main dust-producers in our Galaxy, they do vastly influence their surroundings by releasing kinetic energy and dust in their vicinity. In some cases, they can even trigger star formation.

The circumstellar environment around Blue supergiant stars is often dense and highly anisotropic. Such a break of symmetry has strong implications on the later phases of stellar evolution, as it can affect the stellar angular momentum. This anisotropy could be driven either by fast-rotation and/or by the presence of a companion star.
MATISSE will allow the detection of previously unseen companion stars and the characterization of the gas and dust in the circumstellar environment. 

\keyword{Red supergiant stars}
Besides their blast as supernova, Red supergiants (RSGs) are also contributors to the chemical and dust enrichment of the Galaxy through their intense mass-loss. As RSGs are not experiencing flares nor pulsations, the triggering of their mass loss remains a mystery and may be linked to the magnetic activity, the stellar convection, and the radiation pressure. 

The molecular gas and dust around these stars can be detected best by MATISSE thanks to its wavelength range and angular resolution. MATISSE will provide a full view of the stellar dynamics from the inner part of the photosphere to the outer envelope layers to understand the mass loss mechanism.

\subsection{Lower mass stars: AGB \& pulsating stars}

\keyword{AGB stars}
Most of the low and intermediate mass stars end their lives on the Asymptotic Giant Branch (AGB) phase. During this phase, pulsation and radiation pressure on dust leads to a phase of strong mass loss, during which gas and dust enriched by the products of the stellar nucleosynthesis will be ejected. This mass loss is crucial for the chemical enrichment of the interstellar medium and therefore for the chemical evolution of galaxies. A pivotal aspect of the mass-loss process is its geometry, i.e. the density distribution of the circumstellar envelope of the AGB stars at different scales and different evolutionary phases.

To understand the mass-loss process, it is essential to study the mass-loss from very deep inside the star up to the interface with the interstellar medium. Due to its broad wavelength coverage, MATISSE is a unique instrument to study the different dust and molecular species present in the atmospheres and envelopes of AGB stars. The aim is to achieve a complete view from the upper photosphere to the outer envelope layers and to complement Herschel and MIDI surveys.

\keyword{Envelopes around Cepheids}
Envelopes around Cepheids have been discovered with long-baseline interferometry in the K Band with the VLTI and CHARA 
\cite{2006A&A...448..623K,2006A&A...453..155M}. 
Since then, four Cepheids have been observed in the N band with VISIR and MIDI 
\cite{2009A&A...498..425K,2012A&A...538A..24G,2013A&A...558A.140G}
and one with NACO 
\cite{2011A&A...527A..51G}.
Some evidence has also been found using high-resolution spectroscopy 
\cite{2008A&A...489.1255N,2008A&A...489.1263N}.
The size of the envelope seems to be at least 3 stellar radii and the flux contribution in K band is from 2\% to 10\% of the continuum, for medium- and long-period Cepheids respectively, 
while it amounts to \about\,10\% or more in the N band.
 
MATISSE offers a unique opportunity to study the envelopes of Cepheids. Determining the size of these envelope (as a function of the spectral band) and their geometry (as a function of the pulsation phase), will bring insights in the links between pulsation, mass loss and envelopes. The impact of CSEs on the period-luminosity relation of Cepheids can also be established.

% ----------------------------------------------------------------------------------------------------------------------------
\section{Planetary systems}\label{sec:planetsys}

% Exoplanets
\subsection{Extrasolar planets}
\label{sec:intro}  
\begin{figure} [t]
   \begin{center}
   \begin{tabular}{c} %% tabular useful for creating an array of images 
   \includegraphics[height=6cm]{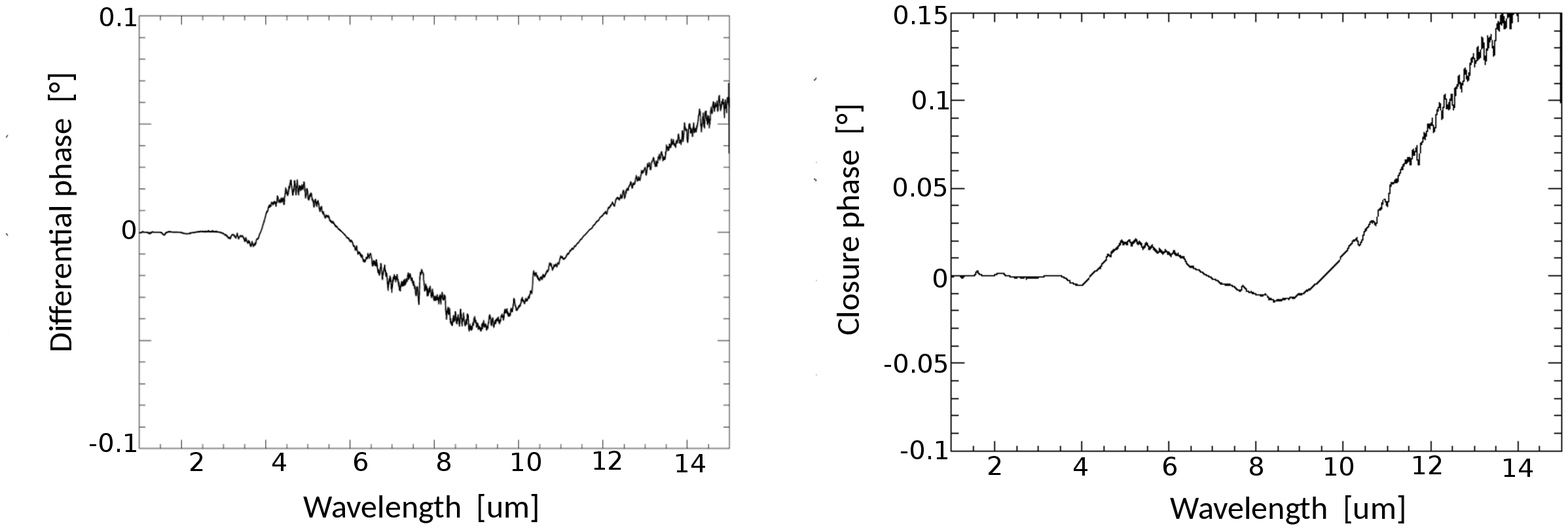}
   \end{tabular}
   \end{center}
   \caption[example] 
%>>>> use \label inside caption to get Fig. number with \ref{}
   { \label{fig:gliese86} 
Left panel: Expected differential phase signal produced by Gliese 86b between 1~$\mu$m and 15~$\mu$m, with a mean projected baseline of about 110~m. Right panel: Expected closure phase signal between 1~$\mu$m and 15~$\mu$m, with the UT1-UT3-UT4 triplet.}
\medskip\medskip\medskip
   \end{figure}

\begin{figure} [t!]
   \begin{center}
   \begin{tabular}{c} %% tabular useful for creating an array of images 
   \includegraphics[height=5.5cm]{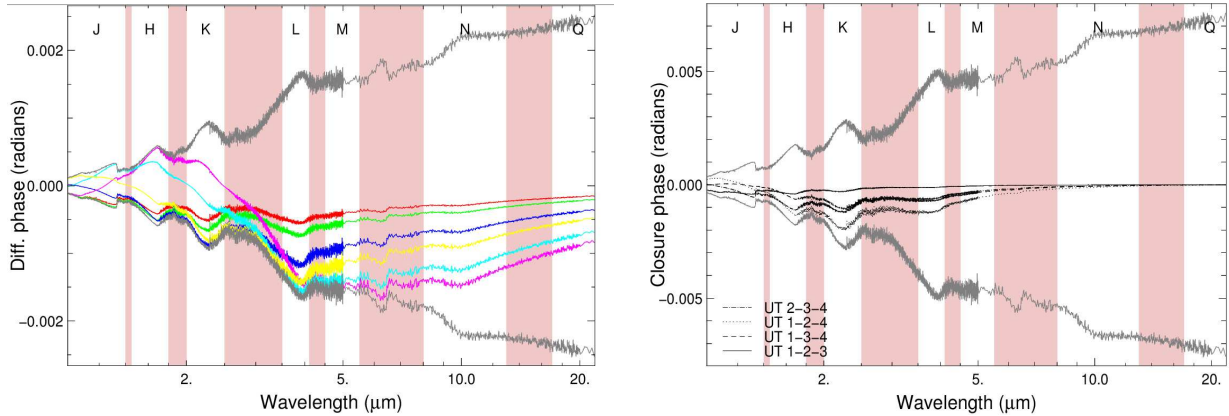}
   \end{tabular}
   \end{center}
   \caption[example] 
%>>>> use \label inside caption to get Fig. number with \ref{}
   { \label{fig:tauboo} 
Expected differential phases (left) and closure phases (right) for $\tau$~Boo b with MATISSE (L/M and N bands).}
   \end{figure}

Since the discovery of 51 Peg b, the presence of Jupiter-mass planetary companions at short ($< 1$~au) orbital distances has raised speculations about their formation and evolution mechanisms.  
The mid-infrared spectral domain is well adapted to the observation of close-in or young Extrasolar Giant Planets (EGPs). Due to their high effective temperature, these sources are significantly luminous in this wavelength domain. Atmospheric composition, planetary mass, and orbit inclination of extrasolar planets around nearby stars may be studied using IR interferometry at the VLTI \cite{2000SPIE.4006..269S,2000SPIE.4006..407L,2006MNRAS.367..825V}. Direct characterization of known EGPs can be imagined with the MATISSE instrument by using differential phase and closure phase.
Assuming $(i,j)$ the pair of telescopes $i$ and $j$, the approximated expression of the differential phase produced by a close-in EGP is:
\begin{equation}
\phi_{ij}(\lambda)\simeq\frac{I_{\rm planet}(\lambda)}{C_*(u_{ij})I_{\rm star}(\lambda)}\:\sin(2\pi u_{ij}\cdot\rho),
\end{equation}
$I_{\rm star}(\lambda)$ and $I_{\rm planet}(\lambda)$ are the monochromatic fluxes of the two components, $\rho$ is the angular separation, $C_*(u_{ij})$ is the modulus of the intrinsic visibility of the stellar component with $u_{ij}=B_{ij}/\lambda$ ($B_{ij}$: baseline between telescopes $i$ and $j$). The associated closure phase on a closed triplet of baselines is:
\begin{equation}
\Phi_{ij}(\lambda)=\sum_{i,j}\phi_{ij}(\lambda).
\end{equation}
Several EGP candidates could be foreseen for observations with MATISSE using differential phase and closure phase observables. Fig.~\ref{fig:gliese86} and Fig.~\ref{fig:tauboo} show the expected differential phases and closure phases for two favourable close-in EGPs detected by radial velocimetry, Gliese 86b ($M \geq$~4~$M_{\rm J}$, 0.11~au semi-major axis) \cite{2000A&A...354...99Q}, and $\tau$~Boo b ($M \simeq$~5.8~$M_{\rm J}$, 0.046~au semi-major axis) \cite{1997ApJ...474L.115B}.

The tentative exoplanet programs on VLTI/AMBER and MIDI paved the way to MATISSE. In particular, we managed to strip down the precision of measurements to a few tenths of milliradian on Gliese 86b \cite{2010A&A...515A..69M}, but did not succeed yet to lower the accuracy enough. In such a context, MATISSE has features which make it particularly attractive for the observation of close-in EGPs :
\begin{itemize}
\item Observations in the L, M and N bands potentially yield the best possible ratio between the flux expected from an EGP and its star (up to $10^{-3}$ in the case of Gliese 86b and $\tau$~Boo b),
\item The error resulting from the still relatively limited background in L band is low enough to get high precision on phase measurements,
\item Observations in the L, M, and N bands are intrinsically less affected by infrastructure perturbations than observations at shorter wavelengths,
\item The use of four telescopes increases the available closure phases by a factor of four and the available differential phases by a factor of two, if compared to the AMBER instrument. 
\end{itemize}
According to Figs.~\ref{fig:gliese86} and ~\ref{fig:tauboo}, one would need a typical accuracy on closure phases and/or differential phases better than $10^{-3}$~rad ($\sim 0.05^{\circ}$). 
From considerations of the fundamental noises only (no systematics), the L band should be the most advantageous wavelength for exoplanets observations. Preliminary estimates of the fundamental error on the MATISSE phase measurements indicates that an accuracy of $10^{-3}$~rad would be achievable in about 10 hours of observation. These estimates will be refined as the instrument will be fully tested. The exoplanet science case is thus very challenging for MATISSE. To reach the necessary accuracies, the most critical aspect will be the fine correction of the instrumental phase and the $2^{\rm nd}$ order chromatic effects due to the atmospheric water vapor \cite{2004SPIE.5491..588T,2010A&A...515A..69M}.\\
To graduate the difficulty and get the complete properties and biases characteristics of the MATISSE instrument at high dynamic range, it is foreseen to observe first single-lined high-contrast binaries and possibly binary brown dwarves. 
%Such a preparatory study will allow to maximize the scientific return on the exoplanet program. 

% Solar system minor bodies
\subsection{Solar system minor bodies}

Information about the physical properties of asteroids and comets, such as sizes, albedos, masses, composition and  surface characteristics, provides essential constraints to models of the formation and collisional evolution of these bodies\cite{Morbidelli2009Icar..204..558M,BottkeAstIV,JohansenAstIV}.
%One particular information that is missing is the original inventory, size distribution, and composition of asteroid. 
In turn, the properties and the history of minor bodies provide constrains to the processes that led to the formation and the evolution of the terrestrial planets in our Solar System. 
Moreover, asteroids are involved in understanding other key issues in Solar System science, such as the delivery of water and organic molecules to Earth, the danger represented by Near Earth Objects and its mitigation \cite{Harris2014astIV}, the role of impacts in affecting Earth's climate and mass extinction events. 
%
%The missing information 
%
Despite the fact that more than 720.000 asteroids are catalogued today, crucial parameters, such as composition and density, are known only for a few bodies. When compared with the densities of meteorites -- a partial sample of the building blocks of asteroids that survive the passage through the Earth's atmosphere -- one can deduce the bulk densities of these bodies. The latter parameter is directly related to the nature of asteroid interiors (e.g. fragmented, monolithic, differentiated)\cite{Consolmagno2008ChEG...68....1C}.

%
%	internal densities
%	mass
%	volume i.e. shape
%	
%	surface properties
%
%what are the main things MATISSE can do
%

\begin{figure} [t!]
   \begin{center}
   \begin{tabular}{c} %% tabular useful for creating an array of images 
   \includegraphics[height=7cm]{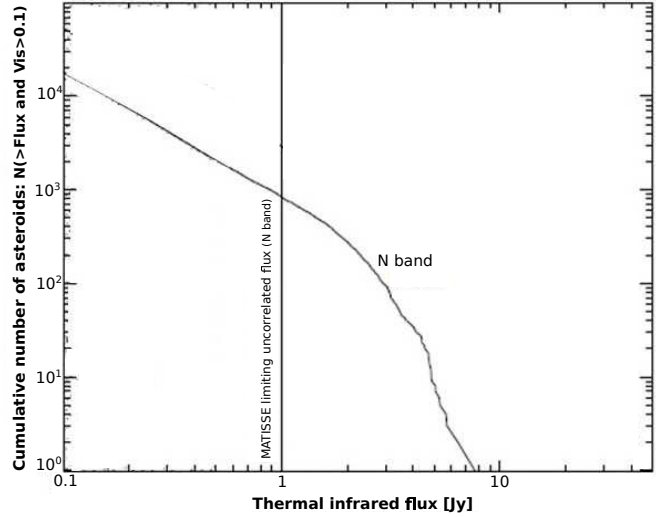}
   \end{tabular}
   \end{center}
   \caption[example] 
%>>>> use \label inside caption to get Fig. number with \ref{}
   { \label{fig:asteroid} 
Cumulative number of Main-Belt asteroids observable with the VLTI in N band as function of their N band flux (visibility $>$ 0.1). Diameters of asteroids are calculated from the known absolute magnitudes and assuming a geometric visible albedo of 0.11. The visibility is calculated assuming a uniform disk and a baseline of 24~m. A vertical line is drawn in correspondence with the sensitivity goal of MATISSE in the N band of 1 Jy (uncorrelated flux).}
   \end{figure}

The VLTI can spatially resolve asteroids in a range of sizes and heliocentric distances that are not accessible to other techniques such as adaptive optics and radar mapping. With temperatures ranging from 200-250 K in the Main Belt to 400-450 K in the near-Earth space, the asteroid emission peaks between 7 and 14 $\mu$m \cite{DelboAstIV}. The MIR domain is thus particularly well suited for the observation of asteroids. The feasibility of MIR interferometric observations of asteroids has been demonstrated with VLTI/MIDI for some Main-Belt Asteroids (MBAs). This leads to characterisation of their physical properties such as size and shape \cite{Delbo2009ApJ...694.1228D}, thermal inertia  \cite{Matter2013Icar..226..419M,Matter2011Icar..215...47M} or internal structure \cite{Carry2015Icar..248..516C}. With the use of four telescopes and an extended spectral coverage down to 3~$\mu$m, MATISSE will allow to go further in the physical characterisation of asteroids and comets. Figure \ref{fig:asteroid} shows that with a limiting flux of 1 Jy in N band, a few hundreds of asteroids become observable with MATISSE. 
   
In the following, two examplary science cases where MATISSE could provide new and decisive constraints are described:

% ----------------------------------------
\keyword{Binary asteroids}
%The determination of the bulk density of an asteroid is a challenging problem requiring knowledge of its mass and volume. The mass is one of the most difficult parameters to measure, and is accurately  (15\% relative error) known for less than 100 asteroids \cite{Carry2012P&SS...73...98C} . There are four robust methods to determine the mass of an asteroid: 
%(1) Asteroid-spacecraft perturbations; 
%(2) Asteroid-asteroid perturbations; 
%(3) Asteroid-planet perturbations; and 
%(4) Observations of the motion of asteroid satellites. The first method is by far the most accurate, but is constrained to the rare instances of a close spacecraft encounter. The second method, tracking the motions of asteroids that gravitationally interact with each other, requires modelling the orbits of multiple asteroids over long periods of time and high accuracy astrometry \cite{Mouret2007A&A...472.1017M,Mouret2008P&SS...56.1819M} . 
%The best data are for the largest asteroids 1 Ceres, 2 Pallas, and 4 Vesta. It is expected that Gaia will enable us to derive the masses of the largest 100 asteroids. 
%For the third method, we note that the largest asteroids can produce measurable perturbations on the motion of Mars. However, like method (2) this technique is limited to the largest asteroids.
A fundamental contribution of MATISSE to the physical characterisation of small bodies will be the determination of the separation of the components, mutual orbital parameters, sizes and shapes of asteroids with satellites. This is by far the most productive method to determine asteroid densities and, in principle, not biased towards large objects. This method can provide accurate masses of asteroids since the orbital period and semimajor axis of the satellite uniquely determine the mass of the system. The best observations yield errors of only a few percent in mass. 

Asteroids with satellites have now been found in every major dynamical group: near-Earth asteroids (NEAs), MBAs, Jupiter Trojans, Centaurs and Trans-Neptunian objects (TNOs). Statistical studies of the occurrence of asteroid satellites indicate that ~15\% of the asteroids with size $<$~20km have a satellite. 
Up to now, kilometer-sized binary asteroids have been studied essentially by radar \cite{Ostro2006Sci...314.1276O} and by the photometric observation of mutual eclipses. 
However, radar is limited to NEAs during favourably close approaches with Earth since the signal-to-noise ratio of radar observations scales with $r^{-4}$, where $r$ is the asteroid-radar distance. In the other dynamical populations, binaries have been discovered through direct imaging using adaptive optics on 8 and 10m class telescopes (VLT, Keck) or using the HST (with its superior resolution in space). However, this technique is limited to well-separated ($>$100 mas) binaries involving large ($>$100 km) bodies. Several of these large bodies have been characterised \cite{Marchis2005Natur.436..822M,Marchis2005Icar..178..450M,Marchis2006Icar..185...39M,Marchis2006Natur.439..565M,Descamps2007Icar..187..482D,Descamps2007Icar..189..362D,Descamps2008Icar..196..578D}. These objects show a remarkable range of orbital and physical characteristics: orbital periods between 0.5 to 80 days, secondary-to-primary size ratios between 0.1 and 1, and an asteroid bulk density between 0.7 and 3.5 g cm$^{-3}$, which highlight the potential of the study of binary asteroids for the analysis of their internal structure.

MATISSE will allow targeting the population of binaries discovered by means of photometric lightcurves. These systems are typically too small to be characterised with direct imaging and too distant for radar. High-angular-resolution observations ($<$100 mas) are thus critical for the characterisation of small and distant binaries in the Main Belt.
The challenge that can be overcome by MATISSE is to determine the semimajor axis of the system. In general this requires spatially resolving the secondary from the primary.  In principle, the density of these bodies can be derived from extensive lightcurve observations over a large range of solar phase angles and several epochs. However, severe difficulties arise in the case of MBAs, which are observable over a relatively moderate range of phase angle \cite{Scheirich2009Icar..200..531S}. The orbital period, however, can be derived accurately by timing of the mutual eclipses.

By combining optical lightcurves and interferometric measurements, we will thus be able to derive full sets of orbital parameters for these systems, including system mass and provide more accurate density estimates. The latter will be derived from the additional knowledge of the size and shape of the primary component, which we will derive directly from the visibility, as done in previous studies \cite{Delbo2009ApJ...694.1228D,Carry2015Icar..248..516C}. A larger program based on this approach can produce asteroid density estimates for bodies in a range of sizes and heliocentric distances that cannot be reached by other means. MATISSE will provide simultaneous observations with four different baselines, leading to six simultaneous visibility points. In the case of binary asteroid studies, this will lead to a much better constraint on the angular separation of the system, and will allow us to determine, in one single observation, the 'true' angular separation as projected onto the plane of the sky (and not only along the baseline direction). 
%An efficient and reliable measurement of the semi-major axis of the system is thus expected.

As already shown\cite{Carry2015Icar..248..516C}, this work will be based on the combination of constraints obtained from optical lightcurves (orbital period, pole orientation, orbital phase) with interferometric measurements of the size and separation of binary components. Such measurement will be performed through model fitting of visibilities using geometrical models. UTs are preferable given the low brightness of the photometric binary targets. The binarity signature in the visibilities will be detectable only if the coherent flux contribution from the secondary component will be higher than the fundamental noises level. High sensitivity in correlated flux will thus be a key element in this study. Both absolute and relative measures will be important for these studies. On one hand, good accuracy on absolute visibility levels will be important to estimate the individual size of the binary components. On the other hand, accuracy on relative measures (i.e., differential visibility) will be a key element to distinguish binarity signature in the visibilities and measure the separation of the components.

% ----------------------------------------
\keyword{Surface properties of asteroids}
Concerning the study of asteroid surface properties, MIDI was already used to derive the value of asteroid thermal inertias\cite{Matter2013Icar..226..419M,Matter2011Icar..215...47M}. The latter is the resistance of a material to temperature change and is a sensitive indicator for the properties of the grainy soil \cite{DelboAstIV} on asteroids. Thermal inertia ($\Gamma$) is defined as $\Gamma = \sqrt{\kappa \rho C}$, where $\kappa$ is the material's thermal conductivity, $\rho$ is its density, and  $C$ is its heat capacity. MATISSE  data will be sensitive to the surface temperature distribution, which is strongly affected by the value of the thermal inertia\cite{DelboAstIV,Durech2014astIV}.

The availability of L \& M bands for MATISSE will allow us to constrain the spatial distribution of water ice and hydrated minerals on the surface of asteroids. This enables us to shed light on the way water has been incorporated and processed through aqueous alteration on the surface of asteroids e.g. by impacts (the water feature is spatially localised on the surface) or by accretion of water/ice rich pebbles (the water feature is evenly distributed over the surface). MATISSE can be used to obtain interferometric observations of asteroids that present hydration and water ice absorption features at $\sim$ 3 $\mu m$. In particular, one can observe asteroids known to present absorption features due to OH vibrational fundamentals ($\sim$2.75 $\mu m$), for Ceres, Pallas and  Hygiea, or due to the first overtone of H$_2$O ($\sim$ 3.1 $\mu m$).
A possible approach will be to measure and to compare the correlated fluxes, visibilities, and closure phases in the 3\,$\mu m$ absorption feature with the ones obtained in the L band continuum which traces the global angular size of the asteroid and possibly sub-structures. As the asteroids will rotate significantly during one single night, one can map the spatial distribution of water ice and hydrated minerals on their surface. In parallel, the N band data can be used to constrain further the surface temperature distribution of these asteroids, and especially their thermal inertia, as successfully done in the past by using MIDI \cite{Matter2013Icar..226..419M,Matter2011Icar..215...47M}.

% Debris disks
%\subsection{Debris disks}
%\todo{tbd: debris disks: Jean-Charles}

% ----------------------------------------------------------------------------------------------------------------------------
\section{Active Galactic Nuclei}\label{sec:agn}

The spectacular angular resolution offered by MATISSE will also provide us with a unique opportunity to study  AGNs. The innermost dust distribution  in AGNs  (the putative {\it torus}) is most likely responsible for the viewing-angle-dependent obscuration of the central engine. Therefore, high spatial resolution studies of the inner dust distribution in AGNs are essential for improving our understanding of AGN unification schemes. Furthermore, the high spectral resolution of 5000 of MATISSE can be used to study the kinematics of the Broad Line Region (BLR).

Previous infrared interferometric observations have shown that the structure of the dust distribution is much more complicated than predicted by classical torus models, because a large amount of dust along the polar direction was discovered in many objects. Therefore, interferometric studies can provide crucial constraints for the innermost mas-scale dust structures in AGNs and and help us to shape a new picture of AGN unification. 

Previous VLTI studies of AGNs in the near- and mid-infrared were challenging for several reasons: (1) The limiting magnitude of AMBER observations of about $K$  $\sim$ 9--10; (2) the limiting H band magnitude of the fringe tracker FINITO of $H$ $\sim$ 6--8; and (3) the two-telescope beam combination of the mid-infrared MIDI instrument, which did not allow us to measure closure phases. In contrast to MIDI, MATISSE will be able to perform aperture-synthesis imaging of AGNs in the L, M, and N bands.

Important objects for MATISSE observations are NGC\,1068, the Circinus AGN, Cen\,A,  the nearest QSO 3C273, and also many fainter objects. Below, we briefly summarize some of the  results obtained already with infrared interferometric observations to illustrate the feasibility of MATISSE  observations of AGNs. These results are also important for the planning of future imaging projects, as artificial substructures may appear in reconstructed images if, for example, the low spatial frequencies of very extended structures are under-sampled due to the lack of short baselines.  In the following paragraphs, we briefly discuss some of the  resolved object structures that will play an important role in  MATISSE  aperture-synthesis imaging.

{\it NGC 1068.} The dust distribution in NGC~1068 was resolved by near-infrared bispectrum speckle interferometry \cite{1998A&A...329L..45W,2004A&A...425...77W}, near-infrared VLTI-VINCI interferometry \cite{2004A&A...418L..39W}, and mid-infrared (8 to 13\,$\mu$m)  VLTI-MIDI interferometry \cite{2004Natur.429...47J,2014A&A...565A..71L}.  The bispectrum speckle interferometry K band image shows a compact, elongated dust distribution with a FWHM Gaussian fit diameter of  $\sim$18$\times$39\,mas or 1.3$\times$2.8\,pc. This structure is extended to the north-west along a position angle of approximately $-$16$^{\circ}$ (the H band image has a  similar structure).  The additionally resolved extended northern component (PA $\sim$ 0$^{\circ}$) has a length of about 400\,mas or 29\,pc.  Mid-infrared MIDI observations \cite{2004Natur.429...47J} resolved  a warm  $\sim$2.1$\times$3.4 pc  structure surrounding a compact ($\sim$0.7\,pc) hot component. Additional MIDI observations \cite{2014A&A...565A..71L} with short baselines allowed a more detailed modeling of the intensity distribution.
 
{\it Circinus.}  To interpret the  large number of obtained mid-infrared MIDI correlated flux spectra and wavelength-differential phases, models consisting of several black-body emitters with a Gaussian brightness distribution and with dust extinction were used\cite{2007A&A...474..837T,2014A&A...563A..82T}. The modeling of all interferometric data reveals that the mid-infrared emission can be explained by three distinct components: A compact, highly inclined (almost edge-on)  disk-like component, a compact circular component,  and an  extended dust structure elongated in polar direction. The disk-like component has a Gaussian FWHM of $\sim$7$\times$35 mas ($\sim$0.14$\times$0.70 pc) and is elongated along a PA of $\sim$44$^{\circ}$, that is almost identical to the PA of the edge-on H$_2$O megamaser disk (PA $\sim$ 44$^{\circ}$) and approximately perpendicular to the PA of the ionization cone (PA $\sim$ 314$^{\circ}$) and the radio lobes (PA  $\sim$  115 and 295$^{\circ}$). The extended component has a size of $\sim$50$\times$80 mas ($\sim$1.0$\times$1.6 pc) and is responsible for 80\% of the mid-infrared emission. It is elongated along a PA $\sim$ 310$^{\circ}$. The derived three-component structure of the dust distribution shows that the dust distribution is more complicated than expected from axisymmetric torus models.

{\it Cen A.} At a distance of only 3.8 Mpc, Centaurus A (NGC 5128) is the nearest radio-loud AGN. VLTI/MIDI measurements \cite{2007A&A...471..453M,2010PASA...27..490B} allowed the  resolution of the  nuclear mid-infrared emission and constrain parameters of the emission components. The MIDI observations can be explained by a two-component model of an unresolved synchrotron point source (diameter $\le$7~mas) and an extended dust emission component.  The extended component is  elongated with a size of about 1.3$\times$0.5 pc \cite{2007A&A...471..453M}. 

{\it  NGC 3783 AMBER observations  in the near- and mid-infrared.}   A K band torus radius of  $\sim$0.74 mas or 0.16  pc was derived from the obtained AMBER visibilities \cite{2012A&A...541L...9W}. Temperature-density gradient models were used to interpret the K band AMBER observations, mid-infrared MIDI observations, and the SED. The results show that an inner hot model component with a temperature of 1400 K and a radius of $\sim$0.16 pc are required in addition to a more extended component in order to explain all observations. This hot component might play a similar role as the puffed-up inner rim discovered in several young stellar objects near the dust sublimation radius.  

{\it MIDI Large Programme.} The {\it MIDI Large AGN Programme} studied the  correlated and total fluxes of 23 AGNs  and derived flux and size estimates at 12 $\mu$m using simple geometric models \cite{2013A&A...558A.149B}. For 18 of the sources, two different nuclear components can be distinguished in the radial plot fits (with different relative flux contributions from the compact and extended components). The different two-component structures probably cause the large scatter of the derived size-luminosity relation.  

{\it AMBER observations of the quasar 3C273 with spectral resolution of 1500.} AMBER observations of the nearest QSO 3C273 ($K$=10) \cite{2012SPIE.8445E..0WP} with spectral resolution of 1500 provided, for the first time, spectrally resolved AGN observations in many spectral channels distributed across the Paschen $\alpha$ line. From the interferograms wavelength-differential phases were derived, which can be used to study the kinematics of the BLR. 

Various torus models have been developed, for example, models with a homogeneous, gravitationally settled dust distribution, clumpy models, and hydrodynamical models\cite{1988ApJ...329..702K,2002ApJ...570L...9N,2005A&A...437..861S,2006A&A...452..459H,2008A&A...482...67S}. MATISSE aperture-synthesis imaging of the nearest AGNs will probably be able to discriminate between these models and constrain model parameters. The observed two-component structure of the dust distribution clearly demonstrates that the dust distribution is  more complicated than has been predicted by classical models. Notably, in several objects, large amounts of dust was discovered along the polar direction. This unexpected complexity clearly requires aperture-synthesis imaging to get a realistic insight in the unknown structure of the dust distribution. Therefore, MATISSE can provide crucial constraints for the innermost dust structures in AGNs. These observations are essential for updating our understanding of AGN unification schemes. Furthermore, high spatial and high spectral resolution interferometric observations will provide a exiting new opportunity to improve our understanding of the kinematics of the AGN BLR.

%1. 1998A&A...329L..45W 
%2. 2004A&A...425...77W 
%3. 2004A&A...418L..39W 
%4. 2004Natur.429...47J 
%5. 2014A&A...565A..71L 
%6. 2007A&A...474..837T 
%7. 2014A&A...563A..82T
%8. 2007A&A...471..453M
%9. 2010PASA...27..490B 
%10. 2012A&A...541L...9W 
%11. 2013A&A...558A.149B 
%12. 2012SPIE.8445E..0WP
%13. 1988ApJ...329..702K
%14. 2002ApJ...570L...9N 
%15. 2005A&A...437..861S
%16. 2006A&A...452..459H
%17. 2008A&A...482...67S

% ----------------------------------------------------------------------------------------------------------------------------
\section{Summary}\label{sec:sum}

We have presented a selection of science cases for MATISSE, the $2^{\rm nd}$ generation mid-infrared spectro-in\-ter\-fe\-ro\-me\-ter for the Very Large Telescope Interferometer.
While this list of science cases is by no means comprehensive, it demonstrates the instrument's great potential for discoveries in a broad range of astrophysical topics. 

Moreover, while MATISSE will provide the basis for the reconstruction of high-angular resolution images in the mid-infrared wavelength range, we have demonstrated that even the direct analysis of the observational data through model fitting will provide far more constraints than it was possible with its predecessor MIDI. 
Furthermore, taking into account the various spectroscopic modes as well as the multi-wavelength aspect, MATISSE has been shown to provide unique insight into the nature of various astrophysical sources, complementary to those that can be derived with long-baseline interferometers and other currently or soon available high angular resolution instruments, operating in the near-infrared and submillimeter/millimeter wavelength range.

%%%%%%%%%%%%%%%%%%%%%%%%%%%%%%%%%%%%%%%%%%%%%% acknowledgements
\acknowledgments 
SW acknowledges  funding through the DFG grant WO 857/13-1.

%%%%%%%%%%%%%%%%%%%%%%%%%%%%%%%%%%%%%%%%%%%%%
% References
\bibliography{rvb_references,mh_refs_michiel,am_exoplanet,md_mypapers,fm_added,gw}   % bibliography data in report.bib
\bibliographystyle{spiejour}   % makes bibtex use spiejour.bst

\end{document}